\documentclass[fleqn,twoside]{article}
\usepackage[headings]{espcrc2}

\readRCS
$Id: espcrc2.tex,v 1.2 2004/02/24 11:22:11 spepping Exp $
\usepackage{graphicx}
\usepackage[figuresright]{rotating}
\usepackage{epsfig}

\title{ILDG Middleware Working Group Status Report}

\author{B. Jo\'o \address[MCSD]{School of Physics, University of Edinburgh, Edinburgh, Scotland, U.K},
        W. Watson \address[MCSD]{Thomas Jefferson National Accelerator Facility, Newport News, Virginia, U.S.A} }
       
\runtitle{Middleware Working Group Status Report}
\runauthor{B. Jo\'o}

\begin{document}
\begin{abstract}
We report on the status of the ILDG Middleware Working Group.
\end{abstract}
\maketitle

\section{INTRODUCTION}
The Middleware Working Group was formed with the aim of designing
standard middleware to allow the interoperation of the data grids of
ILDG member collaborations. Details of the working group are given 
in section \ref{sec:About}.  In this contribution we outline the 
role of middleware in the ILDG, present our proposed middleware
architecture and discuss our current status and future work within
the working group.

\subsection{What is Middleware}
Consider a lattice gauge theory practitioner in the US wishing to
retrieve data from the data grid in the UK. He\footnote{for the sake
of grammar only, we assume that the researcher is male} is used to
the way things work in the US.  However, for historical reasons, the
UK has developed its own {\em drive on the left} grid software before the
advent of the ILDG. The US and UK systems are different. They use
different databases for the catalogues. They use different storage
systems.  How is the researcher to avoid the headache of re-learning
everything he has painstakingly learned about the US data grid?

 The researcher interacts with {\em applications} (such as web
browsers) which we will also refer to as {\em clients}. The actual
databases holding the catalogues and the storage systems are called
the {\em back end}. The {\em middleware} comprises of the layers of
abstractions, interfaces, services and protocols between the
applications and the back end. One piece of middleware for example
would be a front end to a catalogue that an application could send
queries to in a single standardised way, irrespective of the actual
back end database. Another piece of middleware may be an abstraction
such as the concept of a logical name given to a piece of data. The
logical name differs from a filename in the sense that it does not
encode the location of the data. Hence applications dealing with
logical names instead of file names can immediately work between
different grids sharing the name-space of files. However, in order to
retrieve the data, logical names still need to be resolved to
actual file names. This can be done by sending the logical name to a
{\em service} that can return the location details. The file can then
be downloaded using a particular file transfer {\em protocol}, such as
FTP, HTTP, or GSIFTP.

\subsection{Web Services}
Middleware can thus allow the interoperation of data grids given that
the abstractions, protocols, interfaces and services comprising it are
standardised. How can the interfaces and services to be standardised?

In the past, gateways to services were
{\em server} programs which interacted with {\em client} programs
through some messaging protocol. Custom and sometimes unportable
messages were often used. Web services are modern versions of these
server programs, with the difference that the definitions of the
interfaces and the messaging protocol have been standardised. Messages
and interfaces are specified in Extensible Markup Language (XML)
\cite{XML}, using the Simple Object Access Protocol (SOAP) \cite{SOAP}
for messages and Web Services Description Language (WSDL) \cite{WSDL}
for the interfaces. XML, SOAP and WSDL are industry standards defined
and maintained by the W3C consortium \cite{W3C}. 

\section{WEB SERVICE ARCHITECTURE}
\subsection{Overview}
The ILDG middleware architecture is to be based on a collection of stateless web services.  These are to provide a standardised interface to back end services such as a local storage system or the grid service layer of a non ILDG data grid.
\begin{figure}[htb]
\epsfxsize=75mm
\epsffile{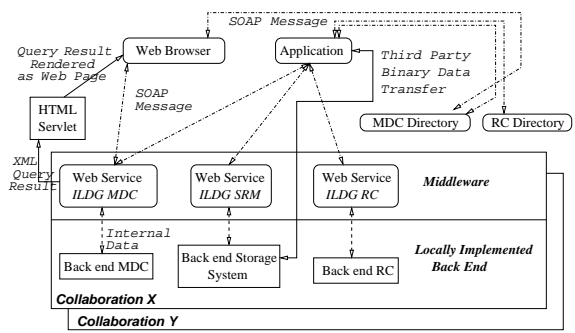}
\caption{The Middleware Web Service Architecture}
\label{fig:MWfig}
\end{figure}

 We illustrate the middleware architecture in figure \ref{fig:MWfig}.
The large rectangles represent the grids belonging to hypothetical
participating collaborations X and Y. These are split into the {\em Middleware}
and {\em Locally Implemented Back End} respectively. The ILDG Middleware
web services appear in the {\em Middleware} half. We have drawn in the three
main web services:
\begin{itemize}
\item
ILDG Metadata Catalogue (MDC),
\item
ILDG Storage Resource Manager (SRM)
\item
ILDG Replica Catalogue (RC) service.
\end{itemize}

These web services interact with back end implementations. The back
end implementations are specific to the participating grid. For
example the back end storage system may be directly controlled by the
ILDG SRM service for collaborations who use SRM to manage their
data. Alternatively, the back end storage system may be some
collaboration specific non--SRM implementation, in which case the ILDG
SRM web service acts as an ``interoperability layer'' allowing ILDG
compliant applications to treat the back end storage as if it was an
SRM. What is important from the point of view of interoperability is
that the applications see a standard web service interface.

Figure \ref{fig:MWfig} also shows possible data transfers between the
application, the web services and the back end. The SOAP messages
exchanged between the application and the web services are shown as
dashed-dotted lines. The solid line between the application and the
back end storage system illustrates the idea of third party transfers,
where the storage system can initiate data upload/download between
itself and the application directly. We also show the hypothetical
case of the ILDG MDC returning the result of a metadata query, and how
this can be rendered from its XML form into a human readable web page
by passing it through an HTML rendering servlet.

Finally we illustrate in figure \ref{fig:MWfig} a possible way for
applications to learn of the existence and whereabouts of the ILDG Web
Services of participating collaborations. This may be done through the
use of {\em directory services}. An application can consult, say, the
MDC Directory service to discover how to contact one or more ILDG MDC
services. It is not yet clear who will operate the directories.
Each collaboration may operate one, or perhaps the ILDG
could maintain one or more global instances of these services, details
of which may be made public on the ILDG web pages.

At this point we should highlight that the middleware working group regards its primary role as that of standardising the web services in the architecture
just presented. The working group does not feel responsible for providing 
the applications or the back end services described.

\subsection{Naming the Data} 
It is envisaged that each item of data in the ILDG will be identified
by a global logical filename (GFN). The space of GFNs will be
partitioned between participating collaborations to avoid name
clashes. The individual collaborations will then be responsible for
managing their own allocated name spaces.

Since the data item may be replicated a GFN may
correspond to several copies of the data. The mapping between GFNs
and individual files is maintained by the RC. We
will refer to these individual filenames as site universal resource
locators (SURLs). SURLs may be presented to the SRM service in
order to retrieve the actual files.

\subsection{Metadata Catalogue}
The ILDG MDC service has the task of allowing the standardised
interrogation of the MDCs of participating collaborations. Applications can
present metadata queries to the ILDG MDC. The replies can be either
GFNs or they can be full or partial metadata instance documents that
are selected by the query. This gives the ILDG MDC {\em read only}
semantics.  It is envisaged that the ILDG MDC will interrogate a
locally implemented MDC which allows for maintenance of the local
catalogue. In other words: insertion/deletion of metadata is expected
to be handled outside of the ILDG framework by the participating
collaborations.

\subsection{Replica Catalogue}
The ILDG RC service has to track various existing copies of a given
file. Essentially, it performs a mapping from a GFN to one or more
SURLs. In order to allow files to migrate it may be necessary for the
back end services or applications to create and remove entries from
the catalogue. Further, we expect that as files migrate, some of the
SURLs returned from the replica catalogue may become invalid. The
burden of dealing with this complexity is pushed onto the
applications. Prototype RCs exist at the Jefferson Laboratory \footnote{in full: Thomas Jefferson National
Accelerator Facility} (JLab) and at Fermilab \footnote{in full: Fermi National Accelerator Laboratory}.

\subsection{The SRM}
The Storage Resource Manager (SRM) has the task of managing the
storage system within a collaborating data grid. SRM is actually a
sophisticated storage resource management system developed between a
variety of institutions. The design of the SRM is lead by the Storage
Resource Management Working Group which is soon expected to become a
Global Grid Forum Grid Storage Management Working Group \cite{SRMWG}.

At the time of writing, version 2.1 of SRM has been defined and the
WSDL definition has been made available online \cite{SRMWG}. The
JLab has an implementation of the SRM 2.1
specification, which was completed in the middle of summer 2004.

The chief envisioned functionality of the SRM in the ILDG, which may
be much more limited than its complete functionality, is to allow the
downloading of files. On the presentation of an SURL, the SRM identifies
the individual file server which holds the file and returns a
transfer URL (TURL) to the file. This is a URL which can be used to
download the actual file, for example by using the {\em wget} utility
with the TURL if the download is to proceed via the HTTP transfer
protocol, or the {\em globus-url-copy} utility if the transfer is to
proceed through GSIFTP.

\section{FUTURE WORK}
We have presented a high level overview of the middleware
architecture as it is currently envisaged. Many details still require
discussion, in particular the detailed WSDL definition of the
interfaces of all the components remains to be completed. This is
particularly true for the case of the MDC and the RC. Work on the MDC
is underway at the CCS at Tsukuba, and at Fermilab. A prototype RC
implementation is nearly ready at the JLab.

The UKQCD collaboration will attempt to implement an SRM compatibility
layer in order to allow the transfer of data between its UK QCDGrid
and a standard SRM such as the one at the JLab.

The interaction of MDCs between grids still needs to be clarified. One
suggestion has been to produce a specification which is {\em
recursive}. In this case one can interact with a root MDC, which
transparently queries lower level MDCs it knows about. Another
suggestion which has already been alluded to is the use of directory
services.

An interesting unresolved issue is how to perform metadata queries. The
metadata working group has proposed a definition of QCDML which assumes a hierarchical XML data model. However hierarchical and XML
databases are not as mature and well known as relational ones,
and it may be desirable to hold the metadata in a relational database
at least at the lowest level. The question is whether to present an
XML (XQuery) or a relational (SQL) view of the databases for the
applications to query, or perhaps to provide both. If the SQL view is to
be provided, it may be useful to define a relational table based
view of the QCDML schema in order to allow straightforward use of SQL back end
databases.

\section{FILE FORMATS AND PACKAGING}
So far we have neglected issues pertaining to the
format and packaging of data. The middleware working group is
currently in discussion with the metadata working group on these
issues.

The key question is whether the ILDG should or should not mandate a
standard gauge configuration file format. Should a file format be
mandated, the question arises as to how to provide tools so that
collaborations can transform the data between the standard format and
their own formats in a straightforward manner.  It is also possible to
not mandate a standard format, but to provide a standard way to
{\em describe the format} of the stored configurations from which tools can
be generated to effect the transformation between formats.

One option for the former case is to maintain program code as part of
the metadata whereas in the latter case the binary data layout may be
completely specified by BinX markup \cite{BinX}. The BinX project also
provides a software library that can be used to perform
transformations on the data such as a rearrangement of indices,
reversal of bit and byte order and the selection of various slices, making it 
straightforward to write programs to transform the data in any desired way. 

\section{CONCLUSIONS}
Currently, the main limitation of the middleware working group is a
lack of manpower. Nonetheless, progress is being made albeit
slowly. The working group plans to hold a working meeting  to be
attended by the conveners and some implementers to finish the
definitions of the web services described here.

\section{ABOUT THE WORKING GROUP} \label{sec:About}
The working group has three joint conveners:
\begin{itemize}
\item
{\em William Watson} from the Thomas Jefferson National Accelerator Facility (JLab), Newport News, VA, USA,
\item
{\em Mitsuhisa Sato} from the Centre for Computational Sciences, Tsukuba, Japan,
\item
{\em B\'alint Jo\'o}  from the School of Physics, University of Edinburgh, Edinburgh, UK.
\end{itemize}
We would also like to highlight the participation of {\em
Eric. H. Neilsen} from Fermilab, Batavia, IL, USA,  who, while not a convener, has been an
extremely active member of the working group contributing a lot of
material including analysis, use cases and samples of WSDL that have
greatly aided the progress of the working group.

Communication in the working group has hitherto proceeded by email
and through the public mailing list \cite{MWList} which is archived at
\cite{MWArch}. Further information about the working group is also
available through the ILDG Web Site \cite{ILDGSite}.

\end{document}